\documentclass[12pt]{iopart}

\usepackage{iopams}  
\usepackage{bm}
\usepackage[final]{graphicx}
\usepackage[square,sort&compress]{natbib}

\newcommand{\eqref}[1]{(\ref{#1})}
\newcommand{\proj}[2]{|#1\rangle\langle#2|}
\newcommand{\openone}[0]{\leavevmode\hbox{\small1\normalsize\kern-.33em1}}

\begin{document}

\eqnobysec

\title{Quantum control of polaron states in semiconductor quantum dots}

\author{Ulrich Hohenester}

\address{Institut f\"ur Physik, Karl--Franzens--Universit\"at Graz,
Universit\"atsplatz 5, 8010 Graz, Austria}
\ead{ulrich.hohenester@uni-graz.at}
\begin{abstract}

We study phonon-assisted dephasing in optically excited semiconductor quantum dots within the frameworks of the independent Boson model and optimal control. Using a realistic description for the quantum dot states and the phonon coupling, we demonstrate that such dephasing has a drastic impact on the coherent optical response. We employ optimal control theory to search for control strategies that allow to fight decoherence, and show that appropriate tailoring of laser pulses allows a complete control of the optical excitation despite the phonon dephasing.

\end{abstract}

\maketitle

\section{Introduction}

In semiconductor quantum dots carriers become confined in all three spatial directions, resulting in a discrete, atomic-like density of states \cite{woggon:97,bimberg:98}. For this reason quantum dots are often referred to as {\em artificial atoms}.\/ Possible applications of quantum dots range from lasers \cite{grundmann:02} over single and entangled photon sources \cite{gerard:99,yamamoto:00,michler:00,benson:00,akopian:06,stevenson:06} to quantum-information devices \cite{barenco:95,loss:98,quiroga:99,imamoglu:99,troiani.prb:00,biolatti:00,troiani.prl:03,chen:00,zrenner:02,li:03,bianucci:04,krenner:05}. However, while for atoms environment couplings can be strongly suppressed by working at ultrahigh vacuum and ultralow temperature, for artificial atoms things are more cumbersome because they are intimately incorporated in the surrounding solid-state environment and suffer from various decoherence channels. This happens even for optical excitation of electron-hole pairs in the states of lowest energy, e.g., exciton or biexction groundstate, causing the deformation of the surrounding lattice (i.e., formation of a {\em polaron}\/ state) whereas relaxation is completely inhibited because of the atomic-like carrier density of states. In coherent optical spectroscopy \cite{shah:96,rossi:02}, which is sensitive to the optically induced coherence, this partial transfer of quantum coherence from the electron-hole state to the lattice degrees of freedom, i.e., phonons, results in {\em dephasing}\/~\cite{borri:01,borri:02b,borri:03}. 

The coupled dot-phonon system is conveniently described within the independent Boson model~\cite{duke:65,mahan:81} and possible generalizations \cite{muljarov:04,muljarov:05}. The independent Boson model is exactly solvable for laser pulses delta-like in time~\cite{krummheuer:02,vagov:02,vagov:03,jacak:03,kruegel:05}, whereas approximate description schemes have to be employed for laser pulses of finite duration. This was first accomplished within a density-matrix approach by F\"orstner et al.~\cite{foerstner:03a,foerstner:03b}, who reported a surprisingly large impact of phonon-assisted dephasing on the coherent optical response. Apparently, this constitutes a serious drawback for quantum-coherence and quantum-information applications in quantum dots, which have recently received considerable interest~\cite{barenco:95,troiani.prb:00,biolatti:00,zrenner:02,li:03,krenner:05,flissikowski:01,flissikowski:04,akimov:06}.

A number of quantum control techniques are known, such as quantum bang-bang control,~\cite{viola:98} decoherence-free subspaces,~\cite{zanardi:97} or spin-echo pulses,~\cite{petta:05} that allow to fight decoherence. However, it is not the system--environment interaction itself that leads to decoherence, but the imprint of the quantum state into the environmental degrees of freedom: the environment measures the quantum system. Optimal control theory~\cite{peirce:88,rabitz:00,borzi.pra:02} allows to design control strategies where quantum systems can be controlled even in presence of such environment couplings without suffering significant decoherence \cite{hohenester.prl:04,hohenester.prb:06}, e.g. by means of laser of voltage pulse shaping. This surprising finding can be attributed to the fact that in the process of decoherence it takes some time for the system to become entangled with its environment. If during this entanglement buildup the system is acted upon by an appropriately designed control, it becomes possible to channel back quantum coherence from the environment to the system.

It is the purpose of this paper to review the quantum control framework developed in refs.~\cite{hohenester.prl:04,hohenester.prb:06} and to apply it to optical control of realistic quantum dots. Contrary to the simple quantum dot model previously used \cite{hohenester.prl:04}, the strong (bi)exciton-phonon coupling will force us to consider not only single- but also multiple-phonon excitations. We have organized our paper as follows. In sec.~\ref{sec:theory} we briefly review some basics of electron-phonon couplings in solids and, more specifically, semiconductor quantum dots, and show how to analytically solve the independent Boson model. Section~\ref{sec:wavefunction} is devoted to a description of our approximate wavefunction scheme. We present results for coherent optical spectroscopy and compare our results with experiment. Finally, we discuss quantum control and optimal quantum control in sec.~\ref{sec:control}, and summarize in sec.~\ref{sec:conclusion}.

\section{Theoretical model}\label{sec:theory}

\subsection{Electron-phonon interaction}

Electrons in solids experience interactions with the lattice degrees of freedom. Let 
\begin{equation}\label{eq:vei}
  \tilde V(\bm r)=\sum_j V_{\rm ei}(\bm r-\bm R_j)
\end{equation}
be the interaction between an electron at position $\bm r$ with the lattice of ions located at positions $\bm R_j$ \cite{mahan:81}. The potential $\sum_j V_{\rm ei}(\bm r-\bm R_j^{(0)})$ for the periodic lattice, with all ions at their equilibrium positions, then gives the bandstructure and the Bloch states of the solid. For small displacements of the ions from $\bm R_j^{(0)}$ one can expand the potential \eqref{eq:vei} in powers of the ionic displacements $\bm Q_j$, to arrive at the electron-phonon coupling
\begin{equation}
  \tilde V(\bm r)=\sum_j \bm Q_j\cdot \nabla V_{\rm ei}(\bm r-\bm R_j)\,.
\end{equation}
This interaction is to be written in terms of operators. For the phonons we adopt the usual displacement operator
\begin{equation}\label{eq:lattice-displacement}
  \bm Q_j=i\sum_{\bm q,\lambda}
    \left(2\rho\omega_{\bm q\lambda}\right)^{-\frac 12}
    e^{i\bm q\cdot\bm R_j^{(0)}}\bm\xi_{\bm q,\lambda}
    \left(a_{\bm q,\lambda}^{\phantom{\dagger}}+a_{-\bm q,\lambda}^\dagger\right)\,,
\end{equation}
with $\rho$ the mass density, $\omega_{\bm q\lambda}$ the frequency of phonon modes with wavevector $\bm q$ and polarization $\lambda$ (longitudinal or transversal), $\bm\xi_{\bm q,\lambda}$ the corresponding polarization vectors, and $a_{\bm q,\lambda}^\dagger$ the bosonic creation operator. We set $\hbar=1$ throughout. For the electrons and holes of the semiconductor material, we expand the charge fluctuations in terms of the fermionic field operators through $c_{\bm k+\bm q}^\dagger c_{\bm k}^{\phantom{\dagger}}$ and $d_{\bm k+\bm q}^\dagger d_{\bm k}^{\phantom{\dagger}}$, respectively. In the following we shall only consider deformation potential coupling to longitudinal acoustic phonons, which is known to be the major dephasing channel for carriers in quantum dots \cite{krummheuer:02}. Interactions with longitudinal optical phonons are known to have a profound impact on the relaxation dynamics in quantum dots \cite{hameau:99,verzelen:00,gladilin:04}, but to be of minor importance for the dephasing dynamics of our present concern owing to the flat optical-phonon dispersion. The phonon interaction then has the form
\begin{equation}\label{eq:hep}
  H_{\rm ep}=\sum_{\bm q}\left(2\rho\omega_{\bm q}\right)^{-\frac 12}
  |\bm q|\left(a_{\bm q}^{\phantom{\dagger}}+a_{-\bm q}^\dagger\right)
  \left(D_e\, c_{\bm k+\bm q}^\dagger c_{\bm k}^{\phantom{\dagger}}-
        D_h\, d_{\bm k+\bm q}^\dagger d_{\bm k}^{\phantom{\dagger}}\right)\,,
\end{equation}
where $D_e$ and $D_h$ are the deformation potentials for electrons and holes, respectively. We have suppressed for conceptual clarity the subscript $\lambda$ in the phonon frequencies and operators. Eq.~\eqref{eq:hep} has the usual form that either a phonon with wavevector $\bm q$ is destroyed and an electron becomes scattered from $\bm k$ to $\bm k+\bm q$, or the electron scatters from $\bm k$ to $\bm k+\bm q$ upon emitting a phonon with wavevector $-\bm q$. The same conclusions apply to holes.

\subsection{Quantum dot states}

Quantum dots are small islands of lower-bandgap material embedded in a surrounding matrix of higher-bandgap material \cite{hawrylak:98,hohenester.review:06}. For properly chosen dot and material parameters, carriers become confined in all three spatial directions within the low-bandgap islands on a typical length scale of tens of nanometers. This three-dimensional confinement results in atomic-like carrier states with discrete energy levels. When the semiconductor is optically excited, an electron is promoted from a valence to a conduction band. In the usual language of semiconductor physics this process is described as the creation of an electron-hole pair or {\em exciton}\/ \cite{haug:93,yu:96}: the electron describes the excitation in the conduction band, and the hole accounts for the properties of the missing electron in the valence band. Electron and hole are conveniently considered as independent particles with different effective masses, which mutually interact through the attractive Coulomb interaction.

\begin{figure}
  \centerline{\includegraphics[width=0.5\columnwidth]{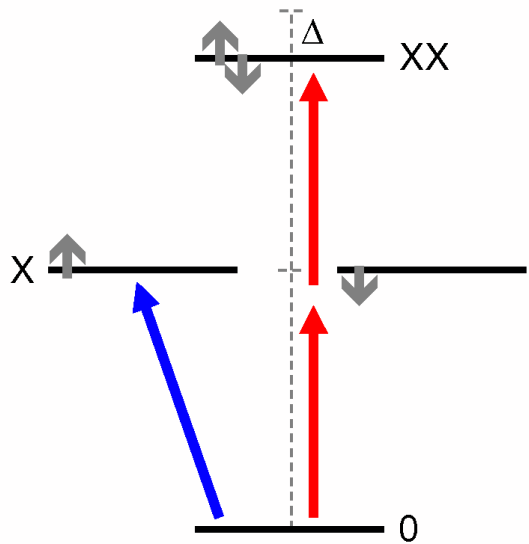}}
\caption{Schematic sketch of the excitonic level scheme in a single quantum dot. Through proper choice of the light polarization, one can promote the dot from the groundstate $0$ to either the spin-up or spin-down exciton state $X$. Within a two-photon process one can also directly excite the biexciton state $X\!X$ of lowest energy, whose energy is reduced by the {\em biexciton binding $\Delta$}\/ with respect to twice the exciton energy \cite{hohenester.review:06}. }\label{fig:sketch}
\end{figure}

In the so-called strong confinement regime the wavefunction extension is comparable or smaller than the excitonic Bohr radius of the bulk material, and the electron-hole states acquire a dominant single-particle character with only slight renormalizations due to Coulomb interactions. This situation approximately corresponds to that of most types of self-assembled quantum dots \cite{woggon:97,bimberg:98,hawrylak:98} where carriers are confined in a region of typical size $10\times 10\times 5$ nm$^3$. To the lowest order of approximation, the groundstate of the interacting electron-hole system is simply given by the product of electron and hole single-particle states of lowest energy $\phi_e(\bm r)$ and $\phi_h(\bm r)$, respectively. Within this work we shall consider the situation where an empty quantum dot is optically excited. By tuning the frequency to the groundstate--exciton frequency and choosing the proper light polarization, fig.~\ref{fig:sketch}, one can selectively excite one of the two spin-degenerate exciton states of lowest energy. The population of excited exciton states can be completely suppressed owing to their large energy separations of several tens of meV. Alternatively, one can also excite within a two-photon process a {\em biexciton}\/ consisting of two electron-hole pairs with opposite spin orientations. In comparison to the exciton energy $\omega_0$, the biexciton energy $2\,\omega_0-\Delta$ is reduced by the biexciton binding $\Delta\sim 2$ meV due to Coulomb correlation effects, which allows to individually address excitons and biexcitons through proper frequency selection. 

In this work we shall investigate optical experiments with short laser pulses, where the semiconductor vacuum is coupled to only the exciton or biexciton groundstate and population of other excitonic states is suppressed because of the large energy splittings. Introducing the usual notation of $0$ for the vacuum and $1$ for the exciton or biexciton state, the system's dynamic in presence of light coupling is governed by the Hamiltonian \cite{hohenester.review:06,walls:95}
\begin{equation}\label{eq:hop}
  H_0=E_0\,\proj 11+
  \left(e^{-i\omega t}\Omega(  t)\,\proj 10+
        e^{ i\omega t}\Omega^*(t)\,\proj 01\right)\,,
\end{equation}        
with $E_0$ the exciton or biexciton groundstate energy, $\omega$ the central frequency of the laser pulse, and $\Omega(t)$ the Rabi frequency associated to the laser pulse envelope. The latter quantity determines the effective light-exciton coupling, and depends on the electric field strength of the laser pulse and the dipole moment of the excitonic transition. In deriving eq.~\eqref{eq:hop} we have made use of the usual rotating-wave approximation \cite{walls:95}. The coupling of the exciton or biexciton groundstate to phonons is governed by the matrix elements $\langle 1|H_{\rm ep}|1\rangle$. Phonon couplings to other excitonic states can be safely neglected because of the large energy separation from the semiconductor groundstate (gap energy $E_g\sim$ eV) and from excited excitonic states (several tens of meV), where typical acoustic-phonon energies are of the order of only a few meV. The total system Hamiltonian thus reads
\begin{equation}\label{eq:indboson}
  H=H_0+\sum_{\bm q}\omega_q\,a_{\bm q}^\dagger a_{\bm q}^{\phantom{\dagger}}+
  \sum_{\bm q} g_{\bm q}\left(a_{\bm q}^{\phantom{\dagger}}+ a_{\bm q}^\dagger\right)\proj 11\,.
\end{equation}
Here $\omega_q=c_\ell\,q$ is the acoustic phonon energy with $c_\ell$ the semiconductor sound velocity, and $g_{\bm q}$ the exciton-phonon matrix element assumed to be real-valued. For the product-type exciton wavefunction $\phi_e(\bm r)\phi_h(\bm r)$ considered above, the matrix element reads \cite{mahan:81,krummheuer:02}
\begin{equation}\label{eq:epmatrix}
  g_{\bm q}=\left(\frac q{2\rho c_\ell}\right)^{\frac 12}
  \left(D_e\int d\bm r\, e^{-i\bm q\cdot\bm r}|\phi_e(\bm r)|^2-
        D_h\int d\bm r\,e^{-i\bm q\cdot\bm r}|\phi_h(\bm r)|^2 \right)\,.
\end{equation}
The integration terms on the right-hand side are precisely the form factors for electrons and holes, respectively. For a biexciton consisting of two electron-hole pairs with opposite spin orientations, the matrix elements \eqref{eq:epmatrix} have to be multiplied by an additional factor of two.

\section{Analytic results}

The system described by the Hamiltonian~\eqref{eq:indboson} is known as the {\em independent Boson}\/ model \cite{mahan:81}. In contrast to the phonon couplings~\eqref{eq:hep}, the independent Boson hamiltonian does not induce transitions between different states. Yet it leads to decoherence. This can be easily seen by writing eq.~\eqref{eq:indboson} in the interaction picture according to the hamiltonian for free phonons, viz.
\begin{equation}\label{eq:indboson.i}
  V(t)=\sum_{\bm q} g_{\bm q}
  \left(e^{-i\omega_q t}\,a_{\bm q}^{\phantom{\dagger}}+
        e^{ i\omega_q t}\,a_{\bm q}^\dagger\right)\proj 1 1\,.
\end{equation}
Through the phonon coupling the quantum dot state becomes entangled with the phonons, where each phonon mode evolves with a different frequency $\omega_q$. If we trace out the phonon degrees of freedom, as discussed in more detail below, the different exponentials $e^{\pm i\omega t}$ interfere destructively, which leads to decoherence. Because this decoherence is not accompanied by relaxation, the process has been given the name {\em pure dephasing}.\/ We shall now study things more thoroughly.
Suppose that an ultrashort laser pulse brings the system at time zero from the groundstate into the superposition state
\begin{equation}\label{eq:superposition}
  \psi=\cos\theta |0\rangle+\sin\theta |1\rangle\,,
\end{equation}
with $\theta$ a small mixing angle. We shall study next how $\psi$ evolves in presence of phonon couplings. In doing so we make the reasonable assumption that the phonons are initially in thermal equilibrium and decoupled from the quantum dot, which will allow us to solve the problem analytically \cite{mahan:81,viola:98}. To understand the essentials of this scheme, it suffices to consider a single phonon mode. The response to the phonon continuum results, as suggested by the term ``independent-Boson model'', by simply summing up the contributions of all phonon modes.

\subsection{Single phonon mode}

Let
\begin{equation}\label{eq:indboson.single}
  h=\epsilon_0\proj 11+\omega\,a^\dagger a+g\left(a+a^\dagger\right)\proj 11
\end{equation}
be the Hamiltonian of a single phonon mode interacting with a two-level system. We next introduce the phonon displacement operator \cite{mandel:95,walls:95,scully:97}
\begin{equation}\label{eq:displacementoperator}
  D(\xi)=\exp(\xi a^\dagger-\xi^* a)\,,
\end{equation}
which has the properties $D^\dagger(\xi) a D(\xi)=a+\xi$ and $D^\dagger(\xi) a^\dagger D(\xi)=a^\dagger+\xi^*$. With the operator $s=\alpha (a^\dagger-a)\proj 11$ one then finds $e^s=\proj 00+D(\alpha)\proj 11$, where $\alpha=g/\omega$. Thus,
\begin{equation}\label{eq:indboson.displaced}
  h=e^{-s}\left(\epsilon_0'\proj 11+\omega\,a^\dagger a\right)e^s\,,
\end{equation}
with $\epsilon_0'=\epsilon_0-g^2/\omega$ the renormalized energy of the two-level system. In other words, the independent Boson hamiltonian \eqref{eq:indboson.single} results from the hamiltonian $h_0=\epsilon_0'\proj 11+\omega\,a^\dagger a$ by shifting through $D(\alpha)$ the phonons when the system is in the upper state, accounting for the fact that the exciton provides a perturbation to the lattice which becomes distorted. This is reminiscent of molecular physics where electronic excitations are accompanied by a change of the binding properties and consequently the molecular structure, though in our case the coupling is much weaker and to a continuum of phonon modes rather than to a few vibronic states.
Consider next the time evolution operator in the interaction picture
\begin{equation}
  u(t,t')=e^{i h_0t}e^{-ih(t-t')}e^{-i h_0t'}=
  e^{i h_0t}e^{-s}e^{-ih_0t}\,e^{i h_0t'}e^s e^{-i h_0t'}\,,
\end{equation}
where we have used eq.~\eqref{eq:indboson.displaced} to arrive at the final expression. The first three terms of the last expression can be simplified to
\begin{equation}
  e^{i h_0t}e^{-s}e^{-ih_0t}=\proj 00+D(-\alpha e^{i\omega t})\proj 11\,,
\end{equation}
and a similar expression follows for the last three terms. Using the property $D(\xi)D(\xi')=D(\xi+\xi')e^{i\Im m\xi\xi'}$ we then arrive at our final expression
\begin{equation}
  u(t,0)=\proj 00+e^{-i\alpha^2 \sin\omega t}
  D(\alpha[1-e^{i\omega t}])\,\proj 11\,,
\end{equation}
where we have set for simplicity $t'=0$.

\subsection{Phonon continuum}

It turns out that a completely similar procedure can be applied in case of a phonon continuum. This is because each phonon mode couples independently to the exciton or biexciton, and the response of the system can be simply computed by summing over different modes. With $s=\sum_{\bm q}\alpha_{\bm q} (a_{\bm q}^\dagger -a_{\bm q}^{\phantom{\dagger}})\proj 11$ and $\alpha_{\bm q}=g_{\bm q}/\omega_q$ one then finds, in analogy to eq.~\eqref{eq:indboson.displaced}, 
\begin{equation}
  H=e^{-s}\biggl(E_0'\proj 11+\sum_{\bm q}\omega_q
  \,a_{\bm q}^\dagger a_{\bm q}^{\phantom{\dagger}}\biggr)e^s\,,
\end{equation}
with $E_0'=E_0-\sum_{\bm q}g_q^2/\omega_q$ the renormalized exciton or biexciton energy. For the time evolution operator one obtains after some straightforward calculation
\begin{equation}\label{eq:time-evolution}
  U(t,0)=\proj 00+\prod_{\bm q}\left( e^{-i\alpha_{\bm q}^2 \sin\omega_q t}
  D_{\bm q}(\alpha_{\bm q}[1-e^{i\omega_q t}])\right)\,\proj 11\,.
\end{equation}

\subsection{Dephasing}

We next consider the situation where initially the quantum dot is in the groundstate and the phonons in thermal equilibrium, and the quantum dot is brought into a superposition state \eqref{eq:superposition} at time zero. Correspondingly, the phonons will react at later time to the perturbation of the excitonic system. We shall only be interested in the quantum dot system and hence trace over the phonon degrees of freedom. To account for this open-system dynamics we employ a density-operator framework. Let $\rho_0=\proj\psi\psi \otimes\rho_{\rm ph}$ be the initial density operator, with $\psi$ the initial superposition state \eqref{eq:superposition} and $\rho_{\rm ph}$ the operator for phonons in thermal equilibrium \cite{mahan:81}. Owing to the electron-phonon coupling \eqref{eq:indboson}, which preserves the exciton number, the lower- and upper-state populations $\rho_{00}(t)=\cos^2\theta$ and $\rho_{11}(t)=\sin^2\theta$ do not change with time. In contrast, the polarization
\begin{equation}\label{eq:polarization}
  \rho_{10}(t)=\langle 1|\mbox{tr}_{\rm ph}\, U(t,0)\rho_0U(0,t)|0\rangle
\end{equation}
does. To compute this polarization function, we use the time evolution operator \eqref{eq:time-evolution} and utilize that $\langle 0|U(0,t)|0\rangle=1$. Then,
\begin{equation}
  \rho_{10}(t)=\sin\theta\cos\theta\,\,
  \mbox{tr}_{\rm ph}\prod_{\bm q}
  e^{-i\alpha_{\bm q}^2\sin\omega_q t}D_{\bm q}(\alpha_{\bm q}[1-e^{i\omega_q t}])\,.
\end{equation}
It can be shown \cite{mahan:81,breuer:02,barnett:97} that in thermal equilibrium the expectation operator of the displacement operator is $\langle D(\xi)\rangle=e^{-|\xi|^2(n+\frac 12)}$, with $n$ the usual Bose-Einstein distribution function for a given phonon mode. Putting together all results and performing some simple manipulations we arrive at our final result \cite{hohenester.review:06}
\begin{equation}\label{eq:correlation}
  \rho_{10}(t)=\sin\theta\cos\theta\,e^{
  -\sum_{\bm q}\left(\frac{g_{\bm q}}{\omega_q}\right)^2\left(
  i\sin\omega_qt+(1-\cos\omega_qt)\coth\frac{\beta\omega_q}2
  \right)}\,,
\end{equation}
with $\beta$ the inverse temperature. Equation~\eqref{eq:correlation} is exact and holds for arbitrary coupling constants $g_{\bm q}$. In evaluating the sum in the exponent one replaces $\sum_{\bm q}\to\int d^3q/(2\pi)^3$ to account for the continuum of phonon states.

\begin{figure}
\centerline{\includegraphics[width=0.75\columnwidth]{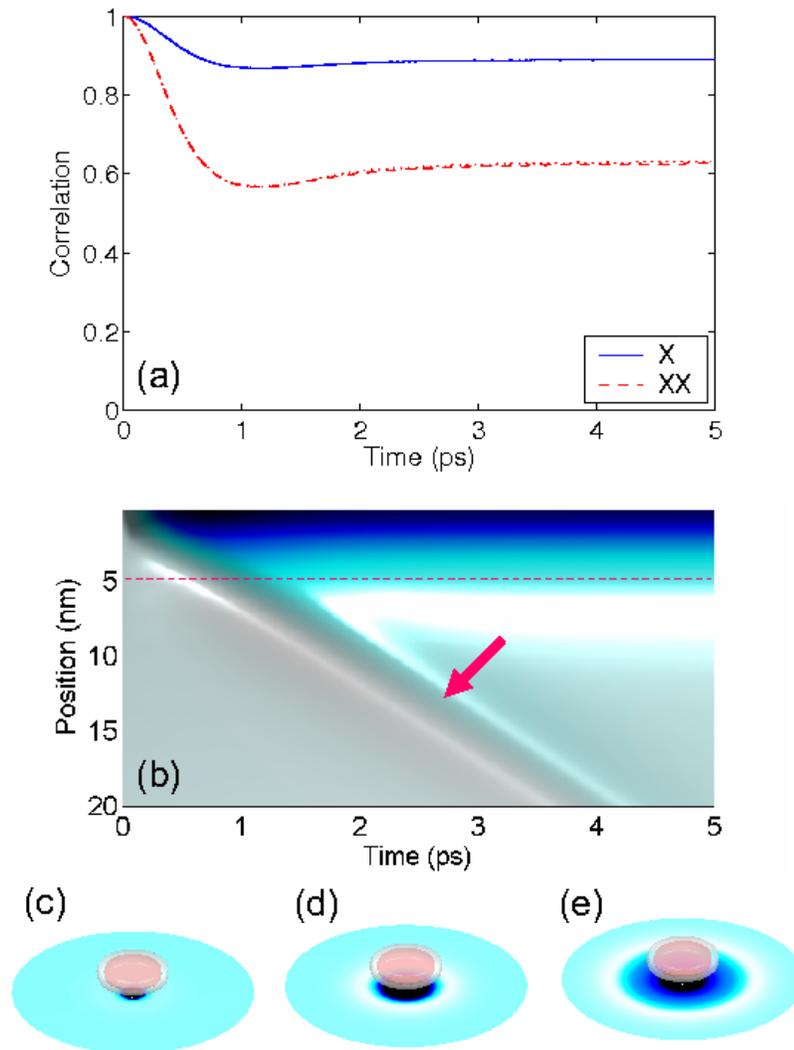}}
\caption{(a) Correlation function \eqref{eq:correlation} for exciton $X$ and biexciton $X\!X$, and for the dot and material parameters listed in table~\ref{table:parameters}. (b) Density plot of the phonon displacement \eqref{eq:lattice-displacement} as a function of time and position, and for the situation where the system is brought at time zero to the superposition state \eqref{eq:superposition}. The dashed line indicates the extension of the exciton wavefunction, and the arrow points to the emitted phonon wave-packet. Panels (c--e) report snapshots of the phonon displacement for selected times of 0.1, 1, and 2 ps.}\label{fig:correlation}
\end{figure}

\begin{table}
\caption{\label{table:parameters}Dot and GaAs-type material parameters used in our calculations. For the electron and hole wavefunctions $\phi_e(\bm r)$ and $\phi_h(\bm r)$ we assume Gaussians with full width of half maximum (FWHM) $L_\ell$ and $L_z$, respectively.}
\footnotesize\rm
\begin{tabular*}{\textwidth}{@{}l*{15}{@{\extracolsep{0pt plus12pt}}l}}
\br
Property & Symbol & Value\\
\mr
Mass density & $\rho$ & 5.37 g/cm$^{-3}$ \\
Sound velocity & $c_\ell$ & 5110 m/s\\
Deformation potential for electrons & $D_e$ & $-14.6$ eV \\
Deformation potential for holes & $D_h$ & $-4.8$ eV \\
in-plane confinement (FWHM) & $L_\ell$ &  8 nm\\
confinement in $z$-direction (FWHM) & $L_z$ &  3.5 nm \\
biexciton binding & $\Delta$ & 2.75 meV \\
\br
\end{tabular*}
\end{table}

Figure \ref{fig:correlation}(a) shows the correlation function $\rho_{10}(t)/(\sin\theta\cos\theta)$ for the realistic dot and material parameters listed in table \ref{table:parameters} and for zero temperature. The correlation function drops on a picosecond timescale to a value of approximately 0.9 and 0.6 for excitons and biexcitons, respectively. This coherence loss is due to phonon couplings, and is stronger for biexcitons than for excitons because of the larger coupling constant. The picosecond timescale is determined by the maximum of the electron-phonon coupling constant \eqref{eq:epmatrix}, which, in turn, is governed by the electron and hole form factors. Thus, for small dots the correlation drops faster than for large dots because of the larger portion of phonon modes to which carriers in small dots can couple. Phonon-assisted dephasing was first observed by Borri and coworkers \cite{borri:01,borri:02b,borri:03} in four-wave mixing experiments, and was confirmed in a large number of independent studies. It constitutes a major obstacle for optical \cite{troiani.prb:00,biolatti:00,troiani.prl:03} and electrical \cite{hohenester.prb:06} quantum-information applications as it spoils the ideal quantum state evolution.

\subsection{Where does the quantum information go?}

As the time evolution described by the operator.~\eqref{eq:time-evolution} is completely unitary, one might wonder how the quantum coherence becomes reduced. In fact, at zero temperature the composite dot-phonon system is described by a single wavefunction. As time goes on the dot degrees of freedom, i.e., exciton or biexciton, become entangled with the phonon degrees of freedom. It is thus the tracing procedure in \eqref{eq:polarization} that provides decoherence of the quantum dot system. Here our ignorance about the environment, i.e., phonon degrees of freedom results in a partial coherence loss \cite{zurek:03}. Where does the quantum information go? To answer this question we plot in panels (c--d) of fig.~\ref{fig:correlation} snapshots of the lattice displacements \eqref{eq:lattice-displacement} subsequent to the preparation of the superposition state at time zero. One observes that it takes some time for the lattice to react to the excitonic perturbation. This is shown even more clearly in panel (b) which reports the time evolution of the lattice distortion along a given direction $x$. Due to the phonon inertia the lattice distortion overshoots, instead of smoothly approaching the new equilibrium position, and a phonon wave packet is emitted from the dots (see arrow) \cite{brandes:05,kruegel:05,hohenester.prb:06} which imprints the quantum information about the superposition state into the environment and thus reduces its quantum properties: the system suffers decoherence. It is thus primarily the abrupt quantum-state preparation that causes excitonic decoherence. We will return to this point later when discussing how such decoherence can be efficiently suppressed by means of quantum control.

\section{Wavefunction approach}\label{sec:wavefunction}

Equation \eqref{eq:correlation} is an exact result that was derived for a delta-like optical excitation. No comparable simple expression can be found for optical excitations with laser pulses of finite temporal width. To account for such situations, one can either approximate the pulse by a comb of delta pulses \cite{vagov:02,vagov:03} or employ a density-matrix description \cite{foerstner:03a,hohenester.prl:04,hohenester.review:06}. In this work we shall use the latter approach. For simplicity we shall restrict ourselves to zero temperature where the composite dot-phonon system can be described by a single wavefunction. The treatment of finite temperatures within a density-matrix approach is similar but more tedious \cite{foerstner:03a,hohenester.prl:04,hohenester.review:06}. Quite generally, our approach is expected to be appropriate also for finite temperatures that are sufficiently small in comparison to those phonon energies $\omega_{q_{\rm max}}$ where the electron-phonon coupling $g_{q_{\rm max}}$ is maximal. For the dot parameters listed in table~\ref{table:parameters} we find $\omega_{q_{\rm max}}\sim 10$ K.

To simplify our analysis, we shall introduce some approximations that are motivated by the analytic expression \eqref{eq:correlation}. We first consider an isotropic quantum dot model with an effective phonon coupling $\bar g_q=\langle q_{\bm q}\rangle_\Omega$ which is obtained by averaging $g_{\bm q}$ over all angles. One can easily check that the correlation function \eqref{eq:correlation} is not altered by this averaging procedure. We next approximate the phonon continuum by a finite set of effective phonon modes. Let $p$ denote an index that labels the modes with wavevectors $q_p=p\,\Delta q$. Here $\Delta q$ is the wavevector spacing. To account for the three-dimensional phase space of phonon modes, we have to modify the effective phonon coupling according to $\tilde g_p=(q_p^2\,\Delta q/(2\pi)^3)^{\frac 12}\,\bar g_{q_p}$ to arrive at our effective Hamiltonian
\begin{equation}
  H=H_0+\sum_p\omega_p\,\tilde a_p^\dagger \tilde a_p^{\phantom{\dagger}}+
  \sum_p \tilde g_p\left(\tilde a_p^{\phantom{\dagger}}+ \tilde a_p^\dagger\right)\proj 11\,,
\end{equation}
with $H_0$ the quantum-dot hamiltonian \eqref{eq:hop} and $\tilde a_p^\dagger$ the creation operator for the effective phonon mode $p$. We next specify our computational Hilbert space. Starting from the quantum dot states $|0\rangle$ and $|1\rangle$ we create one-phonon excitations through $\tilde a_p^\dagger|0\rangle$ and $\tilde a_p^\dagger|1\rangle$, two-phonon excitations through $\tilde a_p^\dagger\tilde a_{p'}^\dagger|0\rangle$ and $\tilde a_p^\dagger\tilde a_{p'}^\dagger|1\rangle$, and so forth. The truncation of this Hilbert space basis is performed on the basis of eq.~\eqref{eq:correlation}, which shows that the importance of phonon modes to the correlation function is governed by $\alpha_{\bm q}=g_{\bm q}/\omega_q$. We correspondingly keep in our Hilbert space only one-phonon excitations with $\tilde\alpha_p>\varepsilon\,\tilde\alpha_{\rm max}$, two-phonon excitations with $\tilde\alpha_p\tilde\alpha_{p'}>\varepsilon\,\tilde\alpha_{\rm max}$, and so forth. Here $\tilde\alpha_p=\tilde g_p/\omega_p$, $\tilde\alpha_{\rm max}$ is the maximum of the $\tilde\alpha$ coefficients, and $\varepsilon$ is a small cutoff parameter. In our calculations we typically keep hundred phonon modes and use a cutoff parameter of $\varepsilon=10^{-2}$.

\subsection{Comparison with analytic results}

The dotted lines in fig.~\ref{fig:correlation} (almost indistinguishable from the solid and dashed lines) show results of our wavefunction simulations for the situation where the dot is initially in the superposition state \eqref{eq:superposition} and the phonons unexcited. One clearly observes that the results of this approximate wavefunction approach is in perfect agreement with those of the analytic expression \eqref{eq:correlation}, thus proving the accuracy of our scheme. We emphasize that the cutoff based on the $\tilde\alpha$ values is crucial to obtain such nice agreement, whereas a mere restriction to one- or two-phonon processes would not work equally well.

\subsection{Rabi oscillations}

\begin{figure}
\centerline{\includegraphics[width=0.85\columnwidth]{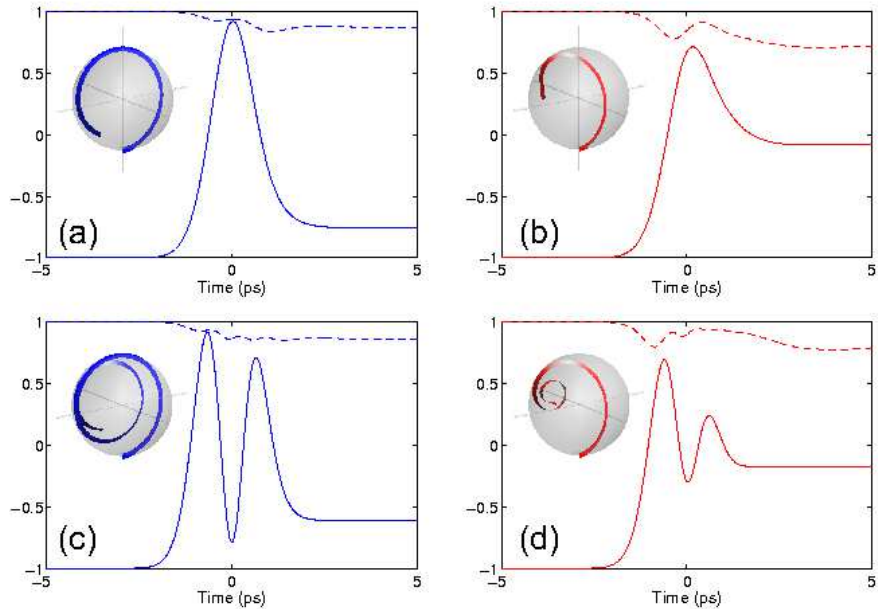}}
\caption{Rabi oscillations in presence of a Gaussian laser pulse for: (a) exciton and pulse area $\theta=2\pi$, (b) biexciton and $\theta=2\pi$, and (c,d) exciton and biexciton for $\theta=4\pi$. The solid line shows the $z$-component of the Bloch vector and the dashed line its modulus. The insets report the Bloch vector trajectories.}\label{fig:rabi}
\end{figure}

We now turn to the situation where a quantum dot initially in the groundstate is excited by a short laser pulse. Throughout we use Gaussian laser pulses with a full width of half maximum of 2.33 ps, approximately corresponding to the experiment of Zrenner et al.~\cite{zrenner:02,krenner:05,stufler:06b}, and assume that the laser frequency is tuned to the renormalized exciton or biexciton transition energy $E_0'$, respectively. Figure~\ref{fig:rabi} shows results of our simulations. We shall find it convenient to analyze our results in terms of the usual Bloch vector $\bm u$ \cite{haug:93,hohenester.review:06}: the $z$ component accounts for the population inversion which is $-1$ when the quantum dot is in the groundstate and $1$ when it is in the excited state; the $x$ and $y$ components account for the real and imaginary parts of the quantum coherence $\rho_{10}$, respectively. The length of the Bloch vector is one for a pure state, and smaller than one when the system has suffered decoherence. 

Panel (a) in fig.~\ref{fig:rabi} shows the Rabi oscillation of an exciton for a laser pulse with area $\theta=\int_{-\infty}^\infty\Omega(t)dt=2\pi$. The inset reports the trajectory of the Bloch vector during the laser excitation. The solid line shows the time evolution of the $z$-component of the Bloch vector, which passes from the groundstate through the excited state (exciton) at time zero back towards the groundstate. However, because of phonon couplings the systems loses part of its coherence to the phonons, and as consequence does not fully return to the groundstate, as it would in absence of phonon couplings. Instead it remains in an excited state: the system has suffered decoherence. This is also apparent from the dashed line that reports the length of the Bloch vector, which decreases from a value of one to approximately 0.9, thus indicating the transition from a pure to a mixed state. Similar behavior is observed in panel (c) for a pulse area $\theta=4\pi$, where the Bloch vector is rotated twice. Finally, panels (b) and (d) show simulations for Rabi floppings of biexcitons, which couple more strongly to the phonons. Correspondingly, the coherence losses are more pronounced. In particular for the $4\pi$ rotation shown in panel (d) the system ends up in a mixed state with equal probabilities for the the upper and lower state populations. We note that these excited exciton or biexciton states would decay on a much longer time scale of nanoseconds due to radiative recombination.

\begin{figure}
\centerline{\includegraphics[width=0.6\columnwidth]{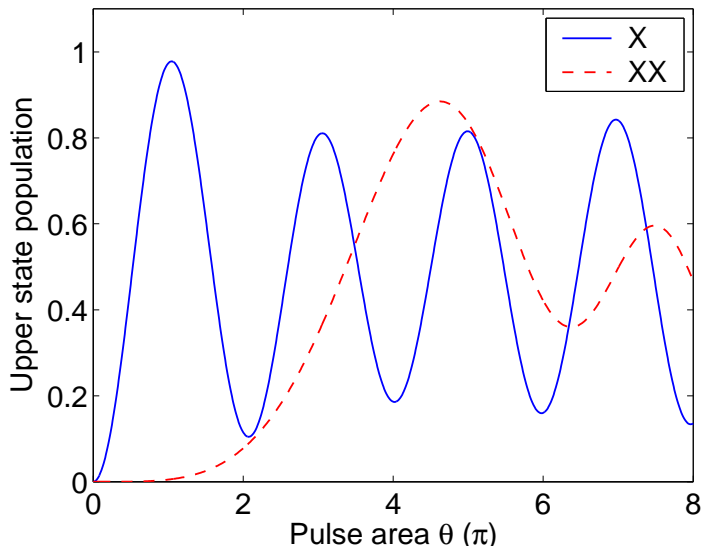}}
\caption{Upper-state populations after completion of Rabi oscillations as a function of pulse area.}\label{fig:rabiarea}
\end{figure}

In fig.~\ref{fig:rabiarea} we plot the upper-state population $\rho_{11}$ as a function of pulse area. For the excitons (solid line) one clearly observes Rabi-type oscillations which are damped for larger pulse areas. Note that for all pulse areas $\theta$ we have used the same pulse widths. Thus, for large $\theta$-values the system oscillates very fast between ground and excited state, and becomes dynamically decoupled from the phonons (see also discussion further below). This is reflected by the fact that the oscillations in fig.~\ref{fig:rabiarea} are less damped for large $\theta$ values. For the biexcitons we use, contrary to the excitation scenario used above, i.e. a Gaussian pulse with given area, a situation that is related more closely to experiment. In fact, in a two-photon experiment the effective Rabi frequency is given by $\Omega_{\rm eff}=\Omega^2/(2\Delta)$ \cite{hohenester.prb:02,stufler:06}, with $\Delta$ the biexciton binding energy (see table~\ref{table:parameters}). The corresponding Rabi oscillations are shown by the dashed line in fig.~\ref{fig:rabiarea}. One observes a different period of the oscillations and a substantially stronger damping in comparison to excitons. We finally emphasize that our results are in almost perfect quantitative agreement with the experimental results of refs.~\cite{krenner:05,stufler:06}.

\section{Quantum control}\label{sec:control}

Above we showed for the independent-boson model that in presence of an exciting laser pulse Rabi flopping occurs but is damped because of phonon decoherence. However, contrary to other decoherence channels in solids where the system's wavefunction acquires an uncontrollable phase through environment coupling, in the independent Boson model the loss of phase coherence is due to the coupling of the electron-hole state to an ensemble of harmonic oscillators which all evolve with a coherent time evolution but different phase. This results in destructive interference and dephasing, and thus spoils the direct applicability of coherent carrier control. On the other hand, the coherent nature of the state-vector evolution suggests that more refined control strategies might allow to suppress dephasing losses.

\subsection{Quantum bang-bang control}

A first step in this direction was undertaken by Viola and Lloyd \cite{viola:98} who considered the situation where a two level system is initially prepared in the superposition state \eqref{eq:superposition} and at later times a sequence of $\pi$-pulses is applied. The operator accounting for $\pi$-rotations is $U_\pi=\proj 01+\proj 10$. Similar to the derivation of the time evolution operator \eqref{eq:time-evolution}, one can show that the time evolution for a sequence of two such $\pi$-rotations is \cite{viola:98}
\begin{eqnarray}
  &&U(t+2\delta t,t+\delta t)U_\pi U(t+\delta t,t)U_\pi\sim\nonumber\\
  &&\quad \proj 00+\proj 11\prod_{\bm q}D_{\bm q}\left(
  -e^{i\omega_qt}\alpha_{\bm q}
  \left(e^{i\omega_q\delta t}-1\right)^2\right)\sim 1+\mathcal{O}(\delta t^2)\,,
\end{eqnarray}
where pure phase factors have been suppressed. The important observation of Viola and Lloyd concerns the fact that for sufficiently small time intervals $\delta t$, i.e., much smaller than the time on which quantum coherence is lost, the system does not suffer decoherence. Rather its dynamics is governed by fast Rabi flopping such that it becomes decoupled from the environment. Consequently, this control strategy has been given the name ``quantum bang-bang control''.

\subsection{Optimal quantum control}

One might wonder whether there exist more simple control strategies that allow to fight decoherence. A general framework for the determination of such control is provided by {\em optimal control theory}\/ (OCT) \cite{peirce:88,rabitz:00,borzi.pra:02,hohenester.review:06}. Here, the objective of the control is quantified through a cost function, which is then minimized subject to the condition that the time dynamics of the coupled dot-phonon system is governed by the Schr\"odinger equation. Optimal control theory is a mathematical device that has found widespread applications for, e.g., molecules \cite{rabitz:00,tesch:02}, atoms \cite{calarco:04}, or semiconductors \cite{hohenester.prl:04,hohenester.prb:06}. 

In the following we shall analyze the situation where a $\pi$-pulse brings the quantum dot from the ground to the exciton or biexciton state. Our control objective is to achieve a maximum upper-state population after completion of the pulse. In the upper state the density operator is of the form $\proj 11\otimes\rho_{\rm ph}$, i.e., the quantum dot is in a pure state. Thus, the transition from the ground to the excited state can only be achieved if decoherence losses are fully suppressed. Let $T$ denote the pulse duration and $\psi(T)$ the terminal wavefunction. The cost function
\begin{equation}\label{eq:costterminal}
  J_T(\psi)=\frac 1 2\langle \psi(T) |\mathbb{P}_0|\psi(T)\rangle
\end{equation}
rates the success of a given control $\Omega(t)$. Here $\mathbb{P}_0=\proj 00\otimes\openone_{\rm ph}$ is the projector on the quantum dot groundstate. $J_T$ is minimal when the system ends in the upper state, irrespective of the final phonon state. To make our optimal-control problem well posed we add an additional term to the cost function, viz.,
\begin{equation}\label{eq:cost}
  J(\psi,\Omega)=J_T(\psi)+\frac\gamma 2\int_0^T \left|\dot\Omega(t)\right|^2\, dt
\end{equation}
that favours control fields $\Omega(t)$ with a smooth time variation. $\gamma$ is a weighting parameter that determines the importance of the two different control strategies of upper-state population and smooth control fields. We shall use small $\gamma$ values throughout, such that the cost function $J(\psi,\Omega)$ is dominated by the first term. The control problem under consideration thus becomes the minimization of the cost function $J(\psi,\Omega)$ subject to the condition that $\psi(t)$ fulfills the Schr\"odinger equation.
Within the field of optimal control theory one uses Lagrange multipliers to turn this constrained minimization problem into an unconstrained one. For this purpose we define the Lagrange function
\begin{equation}\label{eq:lagrange}
  L(\psi,p,\Omega)=J(\psi,\Omega)+
  \Re e\left<p(t),\,i\dot\psi(t)-\left(H+H_{\rm op}(t)\right)\,\psi(t)\right>\,,
\end{equation}
with $H$ the independent Boson hamiltonian \eqref{eq:indboson}, $H_{\rm op}$ the light-matter coupling \eqref{eq:hop}, $\left<u,v\right>=\int_0^T dt\,\langle u|v\rangle$, and $p(t)$ the Lagrange multiplier. We next utilize the fact that the Lagrange function has a saddle point at the minimum of $J(\psi,\Omega)$, i.e. all three derivatives $\delta L/\delta\psi$, $\delta L/\delta p$, and $\delta L/\delta\Omega$ must be zero. Performing usual functional derivatives in eq.~\eqref{eq:lagrange} we obtain after some variational calculation the following optimality system~\cite{borzi.pra:02,hohenester.review:06}
\numparts\label{eq:optimality}
\begin{eqnarray}
&&i\dot\psi=\left(H+H_{\rm op}(t)\right)\psi
\label{eq:oct.forward}\\
&&i\dot p=\left(H+H_{\rm op}(t)\right)p\qquad\label{eq:oct.backward}\\
&&\gamma\ddot\Omega=-\Re e\langle\psi|\Bigl(\proj 01\otimes\openone_{\rm ph}\Bigr)|p\rangle
\label{eq:oct.control}\,,
\end{eqnarray}
\endnumparts
which has to be solved together with the initial and terminal conditions 
\numparts
\begin{eqnarray}
  &&\phantom{i}\psi(0)=\psi_0\label{eq:bc.forward}\\
  &&ip(T)=\mathbb{P}_0\psi(T)\label{eq:bc.backward}\\
  &&\phantom{i}\Omega(0)=\Omega(T)=1\,.\label{eq:bc.control}
\end{eqnarray}
\endnumparts
The right-hand side of eq.~\eqref{eq:bc.backward} follows from the functional derivative $\delta J/\delta\psi$. Notice that while the state equation \eqref{eq:oct.forward} with initial condition $\psi(0)=\psi_0$ evolves forward in time, the adjoint equation \eqref{eq:oct.backward} with terminal condition \eqref{eq:bc.backward} is marching backwards. The control equation \eqref{eq:oct.control} determines the optimal control.
In most cases of interest one is not able to directly guess $\Omega(t)$ such that eqs.~(\ref{eq:oct.forward}--c) are simultaneously fulfilled, and one has to employ an iterative scheme. In this work we follow ref.~\cite{borzi.pra:02} and formulate a numerical algorithm that solves the optimality system (\ref{eq:oct.forward}--c) for a given initial configuration $\psi_0$. To solve this problem, we apply a gradient-type minimization algorithm, which, starting from an initial guess for $\Omega(t)$, determines a search direction for an improved control. We first solve eq.~\eqref{eq:oct.forward} with initial condition $\psi(0)=\psi_0$ forwards in time. Once the wavefunction $\psi(T)$ at time $T$ is computed, the final condition for $p(T)$ can be calculated from eq.~\eqref{eq:bc.backward} and the adjoint equation of motion \eqref{eq:oct.backward} is solved backwards in time. The gradient of $L$ with respect to $\Omega$ becomes
\begin{equation}\label{eq:grad.L}
  \frac{\delta L}{\delta\Omega}=-\gamma\ddot\lambda-
  \Re e\langle\psi|\langle\psi|\Bigl(\proj 01\otimes\openone_{\rm ph}\Bigr)|p\rangle\,,
\end{equation}
which gives the search direction for an improved control that minimizes $J(\psi,\Omega)$. In the following we employ for the minimum search the usual nonlinear conjugate gradient method \cite{press:02}.

\begin{figure}
\centerline{\includegraphics[width=0.85\columnwidth]{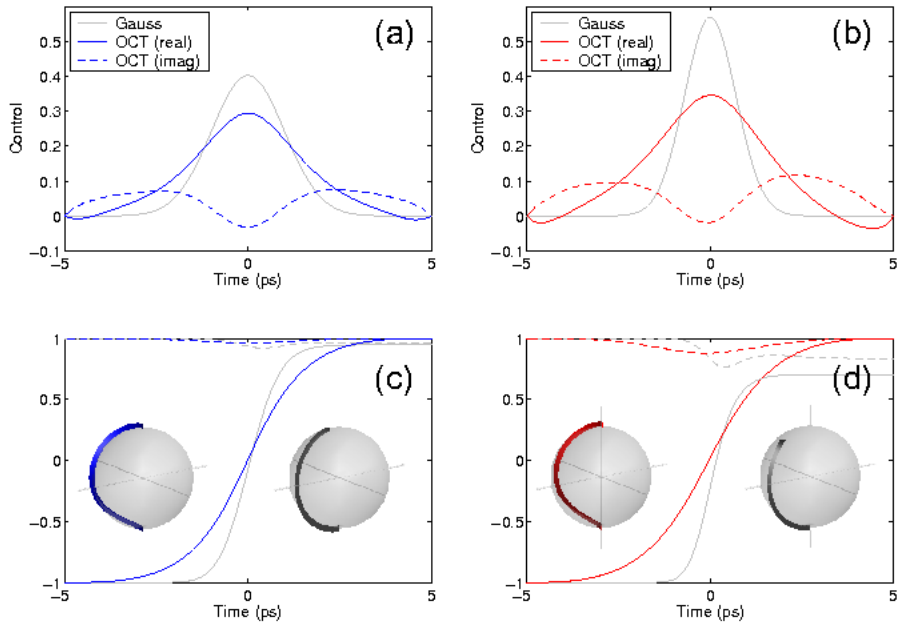}}
\caption{Results of our calculations with a Gaussian $\pi$ (gray lines) and OCT (dark lines) laser pulses. Panel (a) shows $\Omega(t)$ and panel (c) the time evolution of the $z$-component (solid line) and modulus (dashed line) of the Bloch vector for excitons. The insets report the trajectories of the Bloch vector for the OCT (left) and Gaussian (right) pulse. For the Gaussian $\pi$-pulse Rabi flopping occurs but is damped due to electron-phonon interactions. For the optimal control decoherence losses are completely suppressed, and the the system ends up in the desired upper state at time $T$. (b,d) Same as (a,c) but for biexcitons.}\label{fig:oct}
\end{figure}

Results of our optimal-control calculations are shown in fig.~\ref{fig:oct}. Panel (a) reports the Gaussian (gray line) and optimized (dark lines) control fields $\Omega(t)$ for excitons. The corresponding Bloch vector trajectories are depicted in panel (c). We first observe that for the Gaussian control field Rabi flopping occurs, but the system ends up in a mixed state owing to phonon-assisted decoherence. In contrast, the optimized $\Omega_{\rm OCT}(T)$ channels the system from the ground to the excited state without suffering any decoherence losses. {\em Thus, optimal control theory allows to design control strategies that can drastically outperform more simple schemes.}\/ In ref.~\cite{hohenester.prb:06} we showed for a related system that the optimized quantum field strongly suppresses the emission of a phonon wave packet, and allows the lattice to react smoothly to the time varying excitonic configuration. In contrast, for the delta-like control shown in fig.~\ref{fig:correlation} as well as the Gaussian pulse the superposition properties of the quantum dot state are imprinted into the lattice environment by emitting a phonon wave packet, and the system suffers decoherence. Similar conclusions apply for the biexciton control shown in panels (b) and (d). 

Our optimal quantum control strategy differs appreciably from other control strategies. The inherent coupling of electrons to phonons excludes quantum state manipulations in decoherence free subspaces~\cite{zanardi:97} or other quantum-optical control techniques, such as, e.g., stimulated Raman adiabatic passage~\cite{bergmann:98}, where quantum state transfer is achieved through states fully decoupled from the environment. Furthermore, the time dynamics of the phonon degrees of freedom disables spin-echo techniques to restore pure quantum states by means of effective time reversal through $\pi$ pulses. Finally, optimal control only requires smooth laser pulses on the timescale of picoseconds, rather than femtosecond pulses needed for quantum bang-bang control~\cite{viola:98}, where the system has to become dynamically decoupled from the environment. 
We note that an alternative to laser-pulse shaping might be to use laser pulses with fixed, e.g., Gaussian shape, and to perform the quantum control by means of voltage pulses applied to an external gate. Here, the energy detuning between the central laser frequency and the two-level system can be modified~\cite{stufler:06}. Corresponding optimal control calculations indeed demonstrate the applicability of optimized voltage pulses for the purpose of efficient exciton flopping in presence of phonon-assisted dephasing.

\section{Conclusions and summary}\label{sec:conclusion}

In conclusion, we have studied the phonon-induced dephasing dynamics in optically excited semiconductor quantum dots within the frameworks of the independent Boson model and optimal control. We have shown that appropriate tailoring of laser pulses allows to control the dot states without suffering significant dephasing losses. The requirements for such laser-pulse shaping are well within the possibilities of presentday technology. We attribute our finding to the fact that in the process of decoherence it takes some time for the system to become entangled with its environment. If during this entanglement buildup the system is acted upon by an appropriately designed control, it becomes possible to channel back quantum coherence from the environment to the system. We therefore believe that our findings are relevant for a much broader class of solid state systems where quantum information is encoded in long-lived quasi-groundstates with small energy separations, such as electron or nuclear spins, resulting in slow scattering processes that can be manipulated by means of quantum control.

\section*{Acknowledgments}

This work was supported in part by the Austrian Science Fund FWF under projet P18136--N13.

\providecommand{\newblock}{}

\end{document}